# Shape Selection in Diffusive Growth of Colloids and Nanoparticles

Vyacheslav Gorshkov,[a,b] Alexandr Zavalov,[a] and Vladimir Privman[c]

[a]Institute of Physics, National Academy of Sciences, 46 Nauky Avenue, Kiev 680028, Ukraine

[b]National Technical University of Ukraine "KPI,"
37 Peremogy Avenue, Building 7, Kiev-56, 03056 Ukraine

[c]Center for Advanced Materials Processing, Department of Physics,
Clarkson University, Potsdam, NY 13699, USA

## Abstract

We report numerical investigations of a three-dimensional model of diffusive growth of fine particles, the internal structure of which corresponds to different crystal lattices. A growing cluster (particle) is immersed in, and exchanges monomer building blocks with a surrounding medium of diffusing (off-lattice) monomers. On-surface dynamics of the latter is accounted for by allowing, in addition to detachment, monomer motion to the neighboring vacant crystal sites, according to probabilistic rules mimicking local thermalization. The key new feature of our model is the focus on the growth of a single cluster, emerging as a crystalline core, without development of defects that can control large-scale growth modes. This single, defect-free core growth is imposed by the specific dynamical rules assumed. Our results offer a possible explanation of the experimentally observed shape uniformity, i.e., fixed, approximately even-sized proportions, in synthesis of uniform colloids and nanoparticles. We demonstrate the basic principles of well-defined particle shape emergence in such growth. Specifically, several shapes are possible for a given crystal structure. Formation of shapes that follow the crystal symmetry and are uniform, can be a result of the nonequilibrium nature of the growth process. The shape of a growing particle can be controlled by varying the relative rates of kinetic processes, as well as by adjusting the concentration of monomers in the surrounding medium.

---

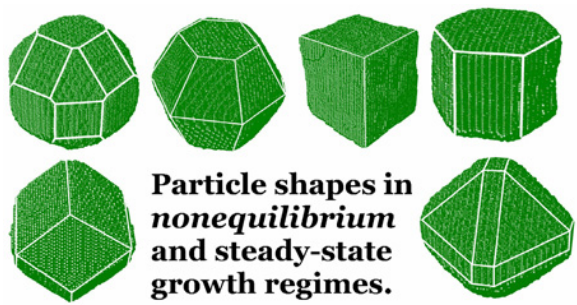

Particle shapes in nonequilibrium and steady-state growth regimes.

Click this area for the published article which will appear in *Langmuir* (2009).

Click this area for *future updates* of this article and for the file with *higher-resolution* images.

# 1. Introduction

Theoretical understanding of mechanisms of growth of "well-defined" (uniform) particles in aqueous and other suspensions, has been an important challenge[1–24] for colloids, and, more recently, for nanoparticles. Uniform colloid particles, of micron and sub-micron sizes, synthesized in solution, have found numerous applications and were extensively studied experimentally, including a large body of recent work,[1–19,25–57] with new developing emphasis on smaller, nanosize particles. Solution synthesis is an important approach because it avoids caging or templating the growing particles, thus allowing for better uniformity in their composition.

Quantitative models that identify diffusional growth mechanisms that can yield narrow particle size distribution in solution, are rather recent. Burst nucleation of nano-crystals, driven by diffusional transport of atom-size building blocks, was shown[2,3] to yield size distributions which are narrow because the smaller-particle side of the distribution, for particle (cluster) sizes below the Gibbs free-energy barrier, is eroded by cluster dissolution, while the larger, over-the-barrier particles grow irreversibly.

The uniformity of colloid particle size has been explained[2,9,18] by a model involving kinetic interplay of burst-nucleation on the nanoscale with further diffusional aggregation of the nanosize primary particles to form polycrystalline colloids. The latter approach yields a good description of the colloid-particle size selection,[1,2,4,7–9,12,16,18] by focusing on the master equations for the secondary particle concentrations for $s$-subunit particles, containing $s$ nanocrystals. The dynamics of this distribution is largely determined by the diffusional transport of building blocks: single- and few-nanocrystal particles, to the growing aggregates.

The aforementioned approaches to uniform size (narrow size distribution) emergence in synthesis of colloids or nanoparticles, while successful in certain regimes, have limitations. Specifically, burst-nucleation per se, only applies for very small clusters. Growth processes of all but the tiniest nanoparticles, involve different and/or additional mechanisms that actually broaden the size distribution[2,3,58] as compared to the predictions of the simplest burst-nucleation model. In synthesis of colloids, the two-stage growth model has been quantitatively successful in explaining[1,2,4,7–9,12,16,18] growth of spherical polycrystalline particles. However, this model's applicability and utility for synthesis of non-spherical particles in not certain. It is likely that such particles are formed by more than a single possible mechanism, and in many cases the process might involve growth of polycrystalline colloids directly from atom-size matter, rather than by the secondary aggregation of "primary-particle" nanosize-crystal building blocks.

While quantitative modeling of particle *size* selection has been partially successful, the challenge of explaining uniformity of the particle *shape* and, more generally, *morphology* in many growth experiments, has remained virtually unanswered. One exception is the "imperfect-oriented attachment" mechanism[6,43,45,59] identified as persistency in successive nanocrystal attachment leading to formation of uniform short chains of aggregated nanoparticles. However, the bulk of the present experimental evidence for the uniformity of particle shapes in solution synthesis, has been collected for particle of colloid-size, from submicron to a couple of microns, and remains largely unexplained quantitatively or even qualitatively.

In fact, colloid particles synthesized and used in applications, can assume a plethora of shapes and structures, depending on their growth conditions. Some particles are grown as single crystals. In many other situations the growth does not yield a well defined size and morphology (structure) distribution at all. However, with properly chosen experimental conditions, there is a large recent body of experimental evidence[1–57,59–60] for growth of well-defined uniform-shape particles. It is the latter growth that will be addressed in the present work.

While in most situations the flat faces of the formed shapes, and the particle proportions, follow the crystal structure of the underlying material, it is experimentally well-established that the particles are almost always polycrystalline, consisting of, and frequently (though not always) growing by aggregation of smaller, nanosize crystal subunits. We point out that in many cases the particles assume shapes that, while uniform, are not those of the standard, equilibrium single-crystal growth. We will refer to well-defined "fixed-shape" particles as those of narrow distribution of shapes and proportions, but not necessarily monocrystalline.

Shape "evenness" is another experimentally observed property, which refers to the tendency of polycrystalline, colloid-size particles to be formed with shapes of fixed and in many cases relatively even proportions. There are, of course, many examples of growth of rods and platelets which are not "even"-shaped. However, they are still uniform in that they usually have similar proportions, as well as surface and internal structure (morphology).

Furthermore, in most cases uniformly shaped particles are also grown with relatively narrow size distributions. However, it is likely that the latter experimentally documented tendency is related to the fact that such particles are of interest in applications and also are easier to characterize. There is no compelling evidence that really narrow size distribution (which, for most applications, would mean no more than about 10% spread in particle modal dimensions) is directly related to particle shape uniformity. Uniform-shape particles can have rather wide size distributions, and it is likely that the mechanisms of size- and shape-selection are not in a one-to-one correspondence.

Indeed, in modeling the *size*, *morphology* and *shape* selection, we have to consider several processes and their competition. Various sets of processes, and their interrelations, will control the resulting particle size, shape, and other structural features. The challenging aspect of the modeling is the large number of processes that compete to yield the final particle structure. In addition to diffusional transport of the atoms (ions, molecules) to form nanocrystals, or that of nanocrystalline building blocks to the particle surface and their attachment to form colloids, these atoms/blocks can detach and reattach. They can also move and roll on the surface, as well as, for nanoparticles as building blocks, restructure and further grow diffusively by capturing solute species.

We know from experimental evidence for (primarily spherical) colloids, for instance, that the arriving nanocrystals eventually get "cemented" in a dense structure, but retain their unique crystalline cores.[2,9,18] The mechanism for this is presently not well understood. In fact, diffusional transport without restructuring would yield a fractal structure.[58,61] On the experimental side, characterization of the various processes occurring on the surface and internally, as time-dependent snapshots during particle growth, has been rather limited, which represents another challenge for modeling, since the results can be only compared to the measured distributions of the final particles, as well as to results of their structural analysis.

In this work, we address aspects of uniform *shape selection* in particle growth in solutions. Numerical difficulties always impose simplifications, and make it impractical to explore all the size-, shape-, and structure-distribution features of growing particles in a single, realistic model. Therefore, approaches are needed that focus on subsets of kinetic processes that control a particular property to be studied. One of the main difficulties in modeling particle shapes numerically,[58,62,63] has been, in fact, the establishment of the crystalline (for nanoparticles) or dense (for colloids) stable "core" on top of which the growth of the structure then continues. In modeling the shape selection, we thus find it practical to focus on the growth of a single (colloid or nano-size) particle as a seed, rather than on a distribution — which would be the main object to consider in studying size selection.

Therefore, we will adopt an approach focusing on modeling a single, defined-core particle growth, with emphasis on the dynamics of its surface and morphological features. The initial formation of the amorphous cluster in nanocrystal growth has been estimated[2,15] to occur for up to roughly order 25 atoms/molecules, followed its restructuring into a crystalline core. The latter can grow to a nanoparticle containing order $10^5$ to $10^8$ atoms. Typically, it takes $10^4$ to $10^7$ nanoparticles to aggregate into colloids. The time scales range from fractions of microseconds for the initial nucleation stages of diffusionally transported atoms, to tens of seconds for the final colloid aggregation.

Our Monte Carlo (MC) approach reported in the present work, was originally designed to address diffusional aggregation of nanocrystals into colloids, but is likely equally suitable to describe aspects of formation of nanoparticles in the diffusional growth stage past their initial burst-nucleation (as well as the growth stage from order $10^2$ to order $10^8$ molecules per crystal in protein crystallization[5,64] after the initial small-cluster formation but before the onset of the really macroscopic growth modes).

We report MC modeling of diffusing "building block" particles/atoms, with the "core" defined by the rule that the particles attaching to the growing, initially small seed, will always "register" with the lattice distances/vectors originally assigned to the seed structure. This approach, while still requiring substantial numerical resources, has the flexibility of allowing to explicitly control the processes of particles (or atoms) "rolling" on the surface and detachment/reattachment: We use thermal-type, (free-)energy-barrier rules. The diffusional transport occurs in the three-dimensional (3D) space, without any lattice. However, the "registered" attachment rule prevents the growing cluster from developing defects and thus, again, ensures the maintenance of a well-defined core. We can then focus on the emergence of the surface and shape morphological features. Our results indicate that there are three regimes of particle growth.

The first regime corresponds to very slow growth rates, for instance, when the concentration of externally diffusion building blocks, to be termed "atoms," is low. In this case, the time scale of the diffusion of already attached atoms on the cluster surface, $\tau_d$, is much smaller than the time scale of the formation of new monolayers, $\tau_{layer}$. Then the shape of the growing cluster is close to the Wulff-construction configuration.[65–67]

The second regime, $\tau_{layer} \ll \tau_d$, corresponds to fast growth and to the development of instabilities of the growing cluster surface. It is expected that the dynamics of the cluster shape is strongly correlated with the spatial density of the diffusional flux, $\vec{\Gamma}(\vec{r})$, near the cluster: $\vec{\Gamma}(\vec{r})$ is maximal near the highest-curvature regions of the surface. As a result, small-scale perturbations of the surface due to random fluctuations, are accompanied by increased diffusional flux of atoms to surface protrusions, which then further grow, provided that near such protrusions the influx of atoms overwhelms the outflow due to on-surface diffusion. Eventually the cluster assumes a form of a clump of sub-structures of smaller sizes.

The third, most interesting regime, corresponds to $\tau_d \sim \tau_{layer}$, with the cluster in a *nonequilibrium* growth mode, but, as demonstrated in this work, it can have an even-shaped form, with well-developed faces that correspond to the underlying crystal structure defined by the seed and by the attachment rules. In this work, we explore the regularities and shape-selection in this nonequilibrium growth regime. Specifically, we consider simple cubic (SC), body-centered cubic (BCC), face-centered cubic (FCC), and hexagonal close-packed (HCP) crystal lattices. For instance, we demonstrate that for the SC case, a cubic particle shape can be only grown in the nonequilibrium growth regime. There are several possible cluster shapes for a given type of crystal symmetry, the realization of each determined by the growth-process parameters.

Our model is simplified, but it apparently captures the key ingredients required for well-defined shape formation in nonequilibrium growth regime: It suppresses formation of "macroscopically" persisting defect structures, as well as, for numerical tractability, it focuses of the growth of a single particle, rather than a distribution. Thus, we believe that it captures the following important feature in both the colloid and nanoparticle growth: In the nonequilibrium growth regime, the model corresponds to the situation when the dynamics of the growing particle's *faces* is *not* controlled by extended defects — which is a well known mechanism that can determine growth modes in nonequilibrium crystallization.[64,68,69] Apparently, this property allows the evolving surface features to overwhelm imperfections, even for colloids that are formed from aggregating nanocrystalline subunits rather than just from the flux of atomic matter onto the surface. The growing cluster faces then evolve to result in well-defined particle shapes and proportions. In fact, we find that the densest-packed, low-index crystal-symmetry faces, which dominate the equilibrium crystal shapes, also emerge in our nonequilibrium regime. However, generally the particle shapes, faces present, and proportions are not the same as in equilibrium.

Obviously, this nonequilibrium growth mode cannot last for very large clusters. However, colloids and nanoparticles are never grown indefinitely. When numerical simulations are carried out for a realistic number of constituent "atoms," our results successfully reproduce many experimentally observed particle shapes. Our numerical approach is detailed in Section 2. Section 3 reports results of our nonequilibrium growth model, as well as auxiliary results for growth close to equilibrium, under steady state conditions to be defined in Sections 2 and 3. We also offer discussion of why we consider the studied nonequilibrium growth regime appropriate for fine-particle synthesis processes. Additional results and considerations, as well as a summarizing discussion, are offered in Section 4.

## 2. Numerical Model

Many of the heuristic assumptions and expectations outlined in the preceding section, are not new. The novel features of our approach are really threefold. First, we focus on growth from a single core, which is initially placed, see Figure 1, near the center of a 3D region, $Q$: Such a nanosize seed is illustrated in Figure 1a. The attachment rules, to be defined shortly, do not allow formation of extended defects. Second, we identify the dynamical processes to include in a simplified model, as well as the appropriate nonequilibrium growth regime and the range of cluster sizes for which the shapes of interest are found. Third, and equally important, to reach theses sizes, the numerical MC simulations had to be carried out for a very large (as compared to reachable in earlier simulations) number of the constituent building blocks (to be termed "atoms"): up to order of a couple of millions. The latter was facilitated by the first two assumptions, especially, our focus on a single seed and proper selection of the dynamical processes. Still, the actual simulations were "large-scale" in terms of modern computational facilities. For example, each cluster-growth realization required between 100 to 400 CPU hours on a 3.2 GHz workstation. In addition, a careful approach to designing the simulation itself was needed, detailed in this section.

Free atoms randomly move in $Q$, and can attach to the growing cluster. The latter thus grows due to diffusional fluxes of free atoms to its surface. The numerical simulations consists of two interrelated parts: modeling the motion of the atoms external to the cluster, and those presently at the surface of the cluster.

### 2.1. *Dynamics of Free Atoms*

In this subsection, we will describe in detail the modeling of the free-atom motion (by diffusion). Actually, this is a rather straightforward part of setting up the modeling approach. However, we offer a detailed presentation for two reasons. First, if not properly programmed, free atom diffusion can be numerically taxing and limiting the efficiency of the simulation. As a result, while the atoms move in a continuum space, modeling their diffusion has to be properly discretized. This leads us to the second reason for carefully accounting for the details: once discretized, we want to be sure that the process of diffusion per se, does not introduce any "lattice" features into the system's dynamics.

Isotropic, off-lattice diffusion of individual atoms can be realized as hopping along randomly oriented unit-length vectors, $\vec{e}_f$, with step-lengths, $l$, randomly taken from a selected distribution. An atom at $\vec{r}_1$ at time $t_1$, will hop to $\vec{r}_2 = \vec{r}_1 + l\vec{e}_f$ at $t_2 = t_1 + \Delta t$. The total number of such hopping events per single cluster-growth simulation in this work, has reached order $10^{12}$. For continuously varying $l$ and $\vec{e}_f$, calculation of the components of the displacement vector $l\vec{e}_f$, involving sines and cosines of directional angles, noticeably affects the overall simulation time. Thus, discretization is warranted. The time steps and displacements are made constant, $\Delta t = \tau$, $l = l_0$ (and sufficiently small). Since the spatial distribution of the atom flux, $\vec{\Gamma}(\vec{r},t)$, near the cluster surface depends on its shape, the hopping length $l_0$ should be comparable to (or smaller than) the lattice spacing in the cluster. In what follows, $l_0$ and $\tau$ are used as units of distance and time. In these units, $l = l_0 = 1$, $\Delta t = \tau = 1$, while the volume $Q$, see Figure 1, was $500 \times 500 \times 500$ in our simulations, and it was aligned with the $x, y, z$ spatial axis.

Similarly, the continuum of $\{\vec{e}_f\}$ was replaced by a discrete set, $\{\vec{e}_f^{(k)}\}$, $k = 1, 2, 3, ..., K$, and the components of the displacement vectors $\{l\vec{e}_f^{(k)}\}$ were pre-computed. Specifically, in the spherical coordinate system the angle $\theta$ with the $z$-axis, was discretized[70] according to $\theta_j = \arccos(1 - (2j-1)/N_\theta)$, with $j = 1, 2, ..., N_\theta$, which mimics the distribution in $0 \leq \theta \leq \pi$, with weight $p(\theta) = (\sin\theta)/2$. The azimuthal angle $\varphi$ was equally spaced in $0 \leq \varphi < 2\pi$, according to $\varphi_i = 2\pi(i-1)/N_\varphi$, with $i = 1, 2, ..., N_\varphi$. In our simulations, we used $N_\theta = 20$, $N_\varphi = 40$. Thus, the set $\{\vec{e}_f^{(k)}\}$ consisted of $K = 800$ different unit-vector directions. However, we also ran several selective simulations with continuum $\{\vec{e}_f\}$, to confirm that the results were unchanged. Such runs were at least several times longer than those with discretized hopping.

Consider a sequence of $M$ vectors $\vec{e}_m$, with $M \gg K = N_\theta N_\varphi$, randomly generated according to the above rules, from the set of values $\{\vec{e}_f^k\}$. For a fixed unit vector $\vec{e}_0$, the mean of the squared projection, $\langle s_1^2 \rangle \equiv \langle (\vec{e}_0 \cdot \vec{e}_m)^2 \rangle = \sum_{m=1}^M (\vec{e}_0 \cdot \vec{e}_m)^2 / M$ should approach the continuum value 1/3. In our simulations, we had $\max\left[\langle s_1^2(\vec{e}_0)\rangle - 1/3\right] = 8 \times 10^{-4}$. Since for fixed hopping length $l = 1$, the quantity $\langle s_1^2 \rangle$ equals the mean-squared displacement along an arbitrary direction, for time step $\Delta t = 1$, then it gives the diffusion constant of the free atoms, $D = \langle s_1^2 \rangle / 2\Delta t = 1/6$, in our dimensionless units.

The boundary of the region $Q$ was defined as the outmost layer of the $500^3$ box, of thickness 3 units. The concentration of free atoms in that layer, $n_0(t)$, was kept constant as a function of time. Specifically, the total number of atoms, $N_b$, in this outer layer of volume $V_b \approx 4.45 \times 10^6$, was monitored, and free atoms were added at random positions (in the outer layer) whenever $N_b$ was found below $N_{b0} = n_0 V_b$. These numbers varied in our simulations, but a representative typical value, used in several runs (see Section 3) was $N_{b0} \approx 8000$, corresponding to density $n_0 \approx 1.8 \times 10^{-3}$, i.e., near the boundary of $Q$ the density was one free atom per cube of linear size $\sim 8$.

Finally, we point out that the time for an atom to diffusively drift the distance $L \approx 200$ from the boundary towards the central region of $Q$, was of order $t_d \approx L^2/(2D) = 1.2 \times 10^5$, and this number also gives the count of the hopping steps involved.

## 2.2. Dynamics at the Cluster Surface

In this subsection, we detail the dynamical rules for atoms which are at the surface of the growing cluster. The initial seed was typically a compact structure, containing up to $\sim 15000$ atoms which were positioned according to the displacement vectors of one of the standard 3D crystalline structures with respect to each other. A key feature of our model was that free atoms adsorbing at the cluster due to diffusion, always attached as the centers of the lattice-structure Wigner-Seitz cells thus continuing to build up the same ideal structure; see Figure 1b. As mentioned, the model focuses on late enough stages of the growth, when the initial seed is already formed. Thus, we avoid the issue of how is the initial seed formed at the initial growth stages. The formation of the initial seed, and its assuming the compact, defect-free crystal structure for nanocrystals, or compactification to a bulk-density polycrystal for colloids, have been discussed in the literature,[2,3,5] but the process is not well understood primarily because of the difficulty with obtaining any relevant experimental data to test theoretical expectations.

For the HCP lattice, we assumed the inter-atomic distance of 1 ($=l$). For the SC and BCC lattices, the cubic lattice spacing, $a$, was taken as $a = l = 1$ in our simulations, whereas for the FCC, we took the cubic lattice spacing $a = 2l = 2$. The diffusive hopping distance, 1, was found small enough for all these cases, to allow the free atoms to explore the lattice cavities (pits) at the outer surface of the cluster. However, the occupied lattice cells were marked so that any attempt by a free atom to hop into their full volumes was rejected, and another hopping direction was then generated for that atom. This prevented atoms from diffusing deep into the cluster. We note that the diffusing free atoms, Section 2.1, were treated instead as point-like and noninteracting. This distinction is in line with our focus on the dynamics of a single central cluster, whereas possible interactions (and thus additional cluster formation) of the free atoms are otherwise ignored.

In most of our simulations, the seed was defined by the lattice cells within a sphere of radius $r_0 = 15$. This choice was made because the "independent particle" approximation is only correct as long as the particles do not meet each other, means, their density is low and they diffuse slowly (which becomes true for large enough particles). In reality, the suspension will have a distribution of multi-atom clusters of various sizes. The processes of cluster-cluster aggregation and of cluster dissolution/breakup, definitely do not favor crystal shapes but rather lead to randomization, i.e., to spherical shapes (as long as we assume that some of the dynamical processes going on also compactify the particle structures). Thus, the whole isolated-particle approximation should be simulated starting only from some large enough sizes below which the clusters are best regarded as spherical.

In our "nonequilibrium" growth simulations the seed atoms were fully immobile. This assumption was made to save run time, based on preliminary-run observations that the seed in such simulations rarely evolved much from its original shape and density. Thus, only the atoms later adsorbed at the seed and the growing cluster, underwent the dynamical motion described in the following paragraphs. However, this rule was not applied to some "steady state" growth runs to be defined in the next subsection. Indeed, if all the arriving atoms after attaching to the seed/cluster, remained immobile, then we would grow fractal structures.[61,71] In reality, surface atoms should be allowed to roll on the growing cluster surface (for instance, in the case of thermal equilibration, to minimize the local free-energy), as well as detach under random agitation. These processes will combine with the cluster growth to yield the formation of a compact particle. In our numerical model, such processes were incorporated as follows.

During the time intervals $\Delta t = \tau = 1$ of the system dynamics, the algorithm of the simulation included analysis of possible moves for each atoms presently attached to the cluster, with the exclusion of those atoms in the original seed and those which are fully surrounded by filled crystal cells. For simplicity, we outline the procedure for the two-dimensional (2D) example shown in Figure 1b. (In reality, our program treated the 3D lattice structures enumerated earlier.) For each surface atom, such as $a$ in Figure 1b, possible dynamical moves were considered.

First, we calculate the probability, $P_{a,mov}$, for atom $a$ to attempt to change its position, defined according to $P_{a,mov} = (p_0)^{k_a}$, where $0 < p_0 < 1$ is a system parameter; $k_a$ is the number of the *nearest-neighbor* cluster atoms (with which atom $a$ "interacts"): $k_a = 2$, in Figure 1b, which corresponds to two out of four maximum possible contacts for the 2D square lattice of our illustrative example. We then generate a random number $r$ ($0 \le r \le 1$). If $r > P_{a,mov}$, then the atom $a$ is left alone and the calculation switches to the next atom (in random order).

If $r \le P_{a,mov}$, then we assumed that the considered atom can attempt hoping to each of its unoccupied *nearest-* and *next-nearest-neighbor* lattice cells (the set of such possible moves was modified in some cases, see Section 3). Specifically, in Figure 1b, atom $a$ can hop to surface cells $b$ or $d$, or detach from the cluster surface, by hopping to cell $c$, or even remain in its original position, cell $a$. In location $b$, atom $a$ will have one "interaction," $k_b = 1$. If hoping to $d$, then $k_d = 2$. Obviously, $k_c = 0$, for hopping to $c$. The probability, $p_\xi$, of each of the possible final positions, here $\xi = a,b,c,d$, was given by the expression of the form $p_\xi = C \exp(\alpha k_\xi)$, with positive parameter $\alpha$, and with the normalization constant $C$ determined via $\sum_{\xi=a,b,c,d} p_\xi = 1$.

If we identify $\alpha = |\varepsilon|/k_B T$, with $k_B$ the Boltzmann constant, then the dependence on a free-energy-like parameter, $\varepsilon < 0$, is suggestive of thermal-equilibration rules for this part of the dynamics. However, we point out that processes of formation of nanoparticles, and more so for colloid-size particles, are in most situations irreversible. Therefore, this "free-energy" designation is questionable. The dynamical rules here rather reflect the expectation that allowed moves of atoms on the surface are still approximately controlled by energy-like considerations (via their binding to the cluster, counted by their number of contacts): the relative probabilities to end up in various possible locations are determined by factors of the form $\lambda^{k_\xi - k_{\xi'}}$. However, the identification of the constant $\lambda \equiv \exp \alpha > 1$ with the Boltzmann weight, and in particular the resulting temperature dependence, $\lambda = \exp(-\varepsilon/k_B T)$, are at best suspect.

Thus, the dynamics should depend on the physical and likely chemical conditions of the cluster and the surrounding suspension, including the temperature. But the latter dependence cannot be assumed equilibrium-like without further experimental or theoretical substantiation. Furthermore, our treatment of the atom-atom/atom-surface interactions generally ignores the possible directionality of the bonding energies, as well as longer-range forces such as double-layer, which will be important to account for in more elaborate treatments.

### 2.3. *Technical Aspects of the Numerical Approach*

In this subsection, our comments apply to the 3D simulations, the *results* of which are detailed in the next section: We provide additional details of our numerical approach. For each 3D simulation, we selected the type of the crystal lattice for the initial seed and thus for the resulting cluster. We also defined the set of lattice vectors, $\{\vec{e}_{int}\}$, along which the atoms interact with each other for the aforementioned determination of the weights $P_{\xi,mov}$. Typically, these will be the nearest-neighbor vectors, e.g., the six (100)-type vectors for the SC lattice. The set of the possible lattice vectors for atom moves, $\{\vec{e}_{mov}\}$, resulting in on-surface displacement or in detachment, was typically larger than the set $\{\vec{e}_{int}\}$, and usually included also the next-nearest neighbors, such as the 12 diagonal lattice vectors of the type (110) for the SC case.

The quantities $p_0$ and $\alpha$, as well as the concentration of the free atoms, $n_0$, in the boundary layer of the volume $Q$, were then selected as the parameters for each simulation run, as were the shape and size of the initial seed, as described earlier. Our numerical "experimentation" suggested that the range of values for the parameter $\alpha$ for which one can seek interesting *nonequilibrium* growth, must be fine-tuned by preliminary *steady-state* runs which are described presently. This steady-state like regime on its own, can also result in interesting particle shapes. However, we argue in Section 3 that, unlike the nonequilibrium regime, this steady-state particle shape evolution is not applicable to practical experimental situations.

Initially, a (spherical) seed of a rather large radius, up to $R \approx 50$, was placed in the center of $Q$. Its evolution with time, was calculated for varying values of $\alpha$, in the absence of any flux of free atoms from the boundary of $Q$, which was made *reflective* for these particular runs. Note that for these preliminary simulations, the atoms in the seed were not fixed, since there were no other atoms for the dynamics. If we took less than a certain cutoff value, $\alpha < \alpha_{cr}$, then the cluster fully disintegrated. However, for $\alpha > \alpha_{cr}$, after some time a steady state developed, with the remaining cluster exchanging atoms with the surrounding dilute solution of atoms, of density two orders of magnitude lower than our aforementioned typical values of $n_0 = 1.8 \times 10^{-3}$ taken for the boundary concentration for our later, nonequilibrium cluster-growth runs. We comment, see Section 3 for details, that the resulting forms of the steady-state clusters, illustrated for the SC lattice in Figures 2 and 3, show some resemblance to the equilibrium (free-)energy-defined configurations. We found that the parameter $p_0$ primarily determined the time scale of approach to the steady state. The value of $\alpha_{cr}$ only weakly depended on $p_0$. The change in the behavior at the "cutoff" $\alpha_{cr}$, is obviously a transition form disintegration to *apparent* steady state behavior on the time scales of the simulations. We do not offer a verdict on whether this is a sharp dynamical phase transition: We simply did not explore the relevant quantities, such as the system- and seed-size dependences of the time scales involved.

For nonequilibrium cluster-growth simulations (with new atoms externally supplied), our numerical "experiments" indicate that nontrivial cluster shapes should be sought (but are not always found) with values of $\alpha$ taken in the range from somewhat less than $\alpha_{cr}$ to up to 30–50% larger than $\alpha_{cr}$. Indeed, for these simulations (which, as described in Section 2.2, bypassed the early-stage spherical-shape growth by starting the large-cluster simulations from the initial seed of radius $r_0 \approx 15$), we found that the cluster would remain nearly spherical and not much larger than the (frozen) seed, due to evaporation of atoms if we take $\alpha$ much smaller than $\alpha_{cr}$. If $\alpha$ is too large then the cluster destabilizes. The SC results are exemplified in Figure 4, which is revisited in Section 3.1. Note that for drawings with more than one particle shape shown, the image sizes in this work were generally not depicted in proportion to the actual linear dimensions, but rather scaled to fit the figure.

The number of atoms in the seed of size $r_0 \approx 15$, is of order $1.5 \times 10^4$. In our cluster-growth simulations, the final clusters typically contained from order $5 \times 10^5$ to $1.5 \times 10^6$ (and sometimes even lager number of) atoms. This means that the seed is overgrown with at least order 30 monolayers of atoms. Initially, the flux of atoms onto the surface of the cluster is maximal, $\Gamma \approx Dn_0/r_0 = 2 \times 10^{-5}$ (recall that we use dimensionless time and distances). Thus, the time of formation of a single monolayer is $\tau_{layer} \approx 1/\Gamma = 5 \times 10^4$, which also counts the number of time-step cycles for dynamical diffusion and on-surface/detachment hopping attempts per each particle, while a monolayer is forming. We note that the number of free atoms in the system is of order $2 \times 10^5$, and "active" surface atoms number several times $10^4$ to several times $10^5$. Multiplication of all these counters indicates the "large scale" nature of the present simulations, typically requiring at least order $10^{12}$ hopping-event calculations per each grown cluster.

Finally, let us address some technical issues related to the cluster structure. While the atoms in the initial seed are frozen, some later-attached atoms, when hopping, can cause pieces (consisting of several atoms) to break off (become detached from) the main cluster. In our simulations, this "connectivity" issue was addressed in two ways. First, if a middle atom in a small "sticking out" structure hopped in such a way that one of more isolated atoms became detached, then these single atoms were made part of the free-atom solution. However, larger "floater" structures were left to evolve according to the system dynamics. Furthermore, a simplified version of the cluster labeling algorithm from percolation theory,[72] was used to classify the cluster connectivity at occasional time instances during the simulation. We found that the "floaters" usually were reattached to the main (seed-connected) cluster and never became large in all but a couple (2–3) of cluster-growth realizations (which were then discarded from our results sample).

We also point out that the algorithm, as formulated, while not allowing the free atoms to diffuse into the cluster structure (as described earlier), does allow trapping internal cavities which will then undergo their own dynamics involving internally trapped pockets of free atoms. We did not attempt a study of the detailed dynamics of such cavities, or their effect on the cluster compactness or density, because we were primarily interested in the emergence of the cluster shape. For the same reason, we did not explore in detail the cluster surface morphology (such as roughness, etc.). All these topics could be a subject of a future study.

It is important to note that *all* the empirical observations summarized in this subsection referred to 3D simulations. Even though we used the 2D case for illustration in Figure 1, no actual 2D simulations were carried out. It is well known that the effect of fluctuations may be more significant in lower dimensions. Therefore, our conclusions cannot be simply "extrapolated" from 3D to 2D without a separate numerical study. For each 3D lattice and parameter values, we actually ran a number of simulations (typically, at least 3) with different random number sequences. The cluster shape (and size) was generally fully reproducible for regimes of interest in our present study, those off relative stable and well defined shapes. In the larger-time or different-parameter-values less stable growth regimes, explored only marginally in the present work, see Sections 3 and 4, the "chaotic" cluster shape does vary significantly from one run to another.

## 3. Results

This section reports results of our simulations, for the selected 3D lattice structures. The most detailed study of the "steady state" regime (reported first), was carried out for the SC lattice, followed by a subsection on the nonequilibrium SC-lattice growth. Results for the other three lattice structures (BCC, FCC, HCP) are reported in later subsections.

### 3.1. *The SC Lattice: Steady State*

In the SC-lattice case, each atom attached to the cluster can have up to 6 bonds with its nearest neighbors, described by the set $\{\vec{e}_{int}\}$ of 6 lattice displacements of the type $(1,0,0)$. The set of displacements/detachments for surface atoms, $\{\vec{e}_{mov}\}$, was defined in two different ways. Case *A*: In this variant, $\{\vec{e}_{mov}\}_A$ included both the set $\{\vec{e}_{int}\}$ and also the 12 next-nearest-neighbor displacements of the type $(1,1,0)$, of length $l_j = \sqrt{2}$. Case *B*: here $\{\vec{e}_{mov}\}_B = \{\vec{e}_{int}\}$. In the latter variant, the dynamics of the surface atoms is slower.

Let us first describe the "steady-state" regime results, as defined earlier, in Section 2.3. The initial large spherical seed in this case was of radius $R = 45$, and its dynamics was simulated without any external supply of free atoms (with reflecting boundary conditions at the outer periphery of the box $Q$ in Figure 1). The shape of the cluster changes in time, as a result of atom evaporation and surface drift. However, with proper selection of the parameters, here exemplified by $p_0 = 0.8$, $\alpha = 3.5$, a steady state develops, of the cluster with the surrounding solution of atoms, for extended time intervals of the simulation. Note that the value of the parameter $\alpha$, defined in Section 2, is here well above our estimates of $\alpha_{cr} \approx 2.7$ for the SC lattice type-*A* dynamics. For the type-*B* variant of the dynamics, we estimated $\alpha_{cr} \approx 2.5$. Atom evaporation and surface drift, result in gradual change of the cluster shape. However, as already mentioned, detached free atoms are reflected at the boundary of the simulation volume $Q$ in this case, and as a consequence, in time a steady state develops and seems to persist for the longest times of our runs.

Figure 2 illustrates the resulting steady-state particle shape for the variant *A* of the SC simulation. We also show a schematic which illustrates that the resulting cluster shape is formed with the type $(100)$, $(110)$, $(111)$ lattice planes, which happen to also be the dense-packed, low-index faces that dominate the low-temperature Wulff construction for the SC lattice.[65-67] However, this superficial similarity with the equilibrium Wulff shape is misleading. Indeed, our system dynamics does not correspond to thermal equilibration. The resulting shape is thus dependent on the dynamical rules. Specifically, Figure 3 shows the shape obtained for the same system but with variant *B* for the displacements/detachments, which imposes a slower surface dynamics.

Evolution of the cluster shape can be facilitated by two mechanisms. The first one is the migration of atoms on the cluster surface. The second process corresponds to preferential detachment of atoms from some parts of the cluster surface and preferential attachment of free atoms at other parts. In order to shed some light on the role of the second mechanism in variant *B* dynamics, we carried out that variant of the calculation with no detachment at all. If a selected surface atom move corresponded to detachment form the main cluster, then that move was not carried out: the atom was left in its original lattice position. In this modified model there were no atoms in the dilute solution. We found that the cluster shape no longer evolved into the one shown in Figure 3. Instead, it reverted to the shape of Figure 2, found for variant *A*. It transpires that the detachment-reattachment mechanism dominates for dynamical rule *B*, as opposed to *A*. We speculate that this is because actually the first mechanism — the on-surface diffusion, is relatively slower for *B* than for *A*, when compared with evaporation/reattachment, for the parameter values and time scales of the simulations.

It is interesting to note that while the evaporation/attachment mechanism has a profound effect on the cluster shape, the actual number of atoms in the dilute solution (quoted in the figure captions for Figures 2 and 3) is surprisingly small as compared to nonequilibrium simulations reported in the next subsection, for which the concentration of free atoms is nearly two orders of magnitude larger, with $\sim 1\text{-}2 \times 10^5$ atoms in the simulation box *Q*.

The above results lead to several interesting observations. First, the particle shape is not universal[73] in the sense expected of many processes that yield macroscopic behavior in Statistical Mechanics: The microscopic details of the dynamical rules do matter. In practical terms this makes it unlikely that particle shapes can be predicted based on generalized arguments such as minimization of some free-energy like quantity. The second conclusion is that the surrounding medium can mediate processes that profoundly affect particle shape. The growth process should thus be considered in a self-consistent formulation that includes the particle's interactions with and the resulting transport of matter to and from its environment.

Another interesting observation is that well defined particle shapes can be obtained in the present steady-state regime. The reader may then ask why not to stop at this point? Why can't this regime be a candidate for predictable (within the present model) and well-defined particle shape selection mechanisms? The answer is in the observed extreme sensitivity to the density of and transport to and form the very dilute surrounding medium. Indeed, in this regime the isolated cluster assumption breaks down: Other clusters (particles) will compete for the "atoms" (solutes) in the dilute solution, and growth mechanisms[74] that involve exchange of matter between clusters will become important (Ostwald ripening). Since such effects are not in the model, we cannot consider it presently predictive for very-dilute-solution (late stage growth) situations.

Let us expand on the observation/speculation that the interrelation between the form of the cluster and the spatial distribution of the flux of free atoms to its surface could provide a early-stage "catalyzing" mechanism for the particle shape selection in the later growth stages. Evaluation of the *steady-state* distribution of the diffusive flux of atoms, $\vec{\Gamma}(\vec{r})$, near an *irreversibly absorbing* surface, is analogous to the problem of calculating the surface electric field, $\vec{E}_{surf}$, at the surface, considered charged and conducting. Obviously, $\vec{E}_{surf}$ is larger in the large-curvature regions of the surface. This means that, generally speaking, protrusions appearing dynamically at the particle surface will be also regions of larger $\vec{\Gamma}(\vec{r})$. This provides a mechanism for instability of the particle surface. Processes such as surface diffusion as well as preferential detachment of atoms, the latter associated with the same regions of larger surface curvature, provide stabilizing mechanisms.[75] Our simulations reported in this subsection, indicate that these three mechanisms combined, suffice to offer growth regimes with, on one hand, the particle shape obviously not thermal-equilibrated, and, on the other hand, stable in that no uncontrollable distortions of the shape arise.

### 3.2. *The SC Lattice: Nonequilibrium Growth*

We now turn to our main, nonequilibrium growth regime, with free atoms at concentration $n_0 \neq 0$ supplied at the outer layer of the simulation box *Q*, etc.; see Section 2. Apparently, the conclusion of the preceding paragraph still applies: interplay of the detachment and attachment of atoms, as well as their surface motion, suffices to yield, in some (but not all) cases well-defined particle shapes. A striking example is presented in Figure 4: growth of a cubic cluster (dynamics variant *A*). We will describe this case in detail presently.

At early times, the flux of atoms is uniform at the (spherical) cluster surface. However, the uniformity of the deposited layer on top of the initial seed is lost rather rapidly, as pointed out in Figure 4a. Atoms deposited at the type-$(100)$ faces, drift on the surface and fill up vacancy sites with larger binding energy at the edges, e.g., atom *1* in Figure 4a. The reversed processes, those decreasing the approximately flat type-$(100)$ faces, occur with smaller probability, e.g., atom *2*. Consideration of 3D configurations in our simulations suggests that the latter processes have to go over a "potential barrier" of a sort, even though there is no equilibration in the regime considered.

Empirical observations also suggest that at later times the cubic shape, Figure 4b, is maintained by interplay of several processes which, in possible continuum descriptions are likely highly nonlinear.[64,68] Specifically, the diffusional fluxes of free atoms indeed favor the edges and corners of the cubic shape. However, an excessive growth of these is prevented by the on-surface drift of atoms toward the centers of the flat faces. The time scale of this drift, $\tau_d$, increases as a square of the characteristic cluster size, $R$, i.e., $\tau_d \sim R^2$, whereas the time scale of formation of new layers, $\tau_{layer}$, grows linearly with the size, $\tau_{layer} \sim R/n_0$ (strictly speaking this result becomes accurate only in the limit of fully adsorbing particles, i.e., no detachment).

If the concentration of free atoms at the boundary of the simulation box is constant, which corresponds to unlimited supply of "building blocks," then the balance between the competing processes of atom arrival (with preference to sharp features of the cluster) and surface drift directed away from these features, can be eventually violated. An unstable regime is then obtained at larger times, with sharp features destabilizing, as illustrated in Figure 4c. This can lead to a "dendritic" instability whereby secondary structures are formed hierarchically, or to destabilization leading to growth of rods, platelets, or other highly uneven shapes. Our simulations were generally not long enough to observe such late-stage growth. However, we give one example of an onset of a rod-shape formation, for a particularly large simulation, in Section 4.

We point out that the regime of plentiful supply of the building block matter, which leads to *fast* "nonequilibrium" growth, is quite common both in nanoparticle nucleation and in colloid growth. However, this available matter: atoms, molecules, ions, sometimes whole nanoparticles in colloid growth, is rapidly depleted. Results such as those presented in Figure 4, suggest that shapes including simple ones (a cube), bounded by low-index planes with symmetry of the material's crystal structure, can be obtained and then practically frozen if the surrounding building-block matter is consumed at an appropriate rate, so that the effective times of fast growth are just right. In many nanoparticle and colloid growth experiments, the appropriate system parameters are frequently determined by empirical trial and error approaches.

We also note that the main difference between the present nonequilibrium regime and the steady-state regime described in Section 3.1, is that the former corresponds to a fast, dominant growth process. Other processes, such as those involving exchange of matter with other clusters, or cluster-cluster aggregation, are simply slower and therefore they do not play a significant role on relevant time scales, as opposed for the situation for the steady-state regime. Thus, shapes found to emerge over intermediate time scales in the fast-growth nonequilibrium regime simulations, are likely to correspond to experimentally realizable situations, though the actual connection between experimental quantities and our MC simulation parameters may not be straightforward to establish.

Let us now comment specifically on the parameter $\alpha$. We already reported our empirical finding that in order to seek well-defined nonequilibrium shapes, the value of $\alpha$ should not be chosen too much below $\alpha_{cr} \approx 2.7$ (all the values presently are for the *A*-variant SC lattice dynamics). Figures 4a,b illustrate the situation with $\alpha = 2.5 < 2.7$: In such cases well-defined shapes can be formed. On the other hand, if $\alpha$ is too small then the growing cluster remains spherical. In nonequilibrium-regime simulations we also observed that if $\alpha$ is increased well over $\alpha_{cr}$, under the conditions of a constant supply of atoms (fixed $n_0$ at the volume boundary), then the particle shape destabilizes. In fact, for the largest values simulated, $\alpha \approx 4$, we observed completely random particle shapes during SC-lattice growth. Thus, there is an approximate range of the $\alpha$ values, $2 < \alpha < 3.5$ for the SC lattice, for which even-shaped, low-index-plane particles are obtained for a range of intermediate times.

Illustrative results for variant *B* are shown in Figure 5, which actually corresponds to the choice $\alpha = 2.5$, which was also our estimate for $\alpha_{cr}$, though we have not explored the issue of the "sharpness" of the definition of the latter quantity form steady-state simulations. There is a transient simulation time interval of small enough *R*, for this dynamics as well, during which the cluster size is in the regime of surface drift ($\tau_d \sim R^2$) proceeding faster that the formation of additional layers ($\tau_{layer} \sim R$). The cluster shape again assumes a form bounded by low-index planes. However, at later times the diffusive flux redistributes to amplify the instability of the vertices of the obtained octahedral shape. The resulting distortion yields a particle with negative surface curvature in some regions; see Figure 5b. For a larger $n_0$, the atom flux is generally larger, and the instability is more pronounced and appears for smaller particle sizes; see Figure 5c, which already shows the onset of the dendritic instability (emergence of secondary structures).

In summary, variation of quantities such as the availability of building block material, i.e., the time dependence of $n_0(t)$, possibly also the building blocks' diffusion constant and some of their on-surface dynamics parameters (if these parameters can be controlled), as well as the attachment/detachment probabilities (which can be modified by varying pH and/or ionic strength), can go a long way towards stopping the process in the well-defined particle shape regime. A careful balance is needed to on one hand achieve the fast-growth nonequilibrium conditions, and, on the other hand not to push the system into the unstable-growth regime.

### 3.3. *The BCC Lattice*

For the BCC lattice simulations, the set $\{\vec{e}_{int}\}$ consisted of 8 nearest-neighbor vectors along the (111) type directions; see Figure 6. However, the surface atom hopping set $\{\vec{e}_{mov}\}$ included not only these 8 vectors $\{\vec{e}_{int}\}$ but also 6 additional type (100) next-nearest-neighbor vectors. (The vectors, etc., are defined in terms of the Cartesian cubic cell, which was of linear dimensions 1 in the hopping length units.)

The BCC lattice, Figure 6, has 12 planes of the (110) type and 6 planes of the (100) type, which will dominate the equilibrium Wulff construction at low temperatures.[65–68] Assuming that all have equal interfacial free energies, the Wulff construction for these faces is shown in Figure 7a. However, even for our steady-state simulations, for the BCC lattice we never obtained particles of these proportions, even though the faces just identified did dominate the growing cluster shapes.

A steady state ($n_0 = 0$, reflecting boundaries, large seed with non-fixed atoms, etc.) simulation result is summarized in Figure 7b, and it demonstrates that the (100) type faces dominate. It might be related to the property that free atoms attaching "next-layer" to such *fully filled* faces are bound to 4 neighbors, whereas for the (110) type faces the number of neighbors is 2. On the other hand, evaporation of an atom from inside a fully filled (100) face requires breaking 6 bonds, whereas from (110), only 4 bonds. Thus, kinetically these faces substantially differ in various energy values involved (in dynamical moves).

An interesting phenomenon encountered for this lattice is that $\alpha_{cr}$, determined as described in the introduction, for the initial seed radius 45, etc., was estimated as $\alpha_{cr} \approx 1.6$. This was the value for which the cluster did not rapidly disintegrate. However, it only assumed well defined *shapes* for larger $\alpha$ values. We used $\alpha = 2$ for Figure 7b and for the nonequilibrium simulations below. Even for $\alpha = 1.8$, the steady-state simulation cluster shape did not have well defined flat faces, though traces of the emerging structure depicted in Figure 7b could be semi-guessed once one knew the larger-$\alpha$ results.

The noticed nonequivalence of the dominant low-index planes, persists in nonequilibrium simulations as well, and can lead to a variety of particle shapes obtained at intermediate time scales. These are illustrated in Figure 8. Specifically, the initially spherical seed first evolves into a rhombic dodecahedron shape; see Figure 8a. Apparently, the formation of the pyramidal vertexes of this shape is driven by the increased diffusional flux to sharp features. However, at later times the overall diffusional flux decreases; recall that $\langle \Gamma(x, y, z) \rangle \sim 1/R(t)$. This results in the vertexes of the rhombic dodecahedron flattening out due to enhanced detachment and diffusion of atoms away from these sharp features: see Figures 8b,c, and the shape shows the tendency to evolve towards the one found in the steady-state simulation reported earlier. In fact, the shape in Figure 8c is actually even closer to the hypothetical shape in Figure 7a, not actually realized in steady-state simulations, but of course the shape in Figure 8c is not obtained in any thermal equilibrium or in a steady state.

The shape at these intermediate times, can be varied by adjusting the system parameters, as illustrated in Figure 8d: nearly a cube, and Figure 8e: a (surviving form shorter times) rhombic dodecahedron. Note that for the parameter values of Figure 8e, the diffusional flux is increased enough not only to maintain the vertexes of the shape on the time scales of our simulation, but to actually show the onset of the destabilization: the piling up of attached atoms, marked by arrows in the figure, similar to that found earlier for a different lattice (and different shape); see Figure 5.

### 3.4. *The FCC Lattice*

The FCC-lattice vectors, etc., will be defined in terms of the Cartesian cubic cell, of linear dimensions 2 in our hopping-length units. Still, the (111) direction for instance, refers to the vector with these cubic-cell coordinates, and collectively to planes perpendicular to it and to similar lattice vectors in other direction. This notation is self-explanatory, and we do not treat the lattice structure in the formal notation of crystallography, nor do we use the Miller indices for lattice plane designation. We note that for the FCC case, $\alpha_{cr} \approx 0.9$.

For the FCC lattice, the densely-packed, low-index faces that tend to be present[65–68] in equilibrium shapes, are (111), as well as (100) and (110). It is therefore natural to expect (some of) these to also show up in our simulations. While each surface atom at such faces on average has binding energy $6\varepsilon$, for dynamical processes we also have to consider the binding of attaching atoms on top of these faces. The fist atom settling on (100) and starting a new layer, has binding energy $4\varepsilon$, and for (111), we get $3\varepsilon$.

The corrugated (ridged) for FCC, (110) type faces were not seen in our nonequilibrium simulations, reviewed shortly, even though the first extra-atom binding energy is $5\varepsilon$. It seems that each fragment of such an over-layer, once formed, acts as a sink for fast additional atom attachment and little detachment, thus becoming the base for a growing wedge bounded by (111) type faces. A posteriori this seems to be a general mechanism, encountered earlier for the SC lattice: The binding energy of an atom at a type-(110) SC face, is larger than that at a type (100) face. For nonequilibrium growth, the (110) faces are not obtained, but rather provide "bases" for emerging wedges involving (100) type faces, resulting in a cube: Figure 4b, instead of the shapes shown, for instance, in Figures 2 or 3. In fact, the expectation that the slowest growing faces are the surviving ones, is quite accepted in crystal growth literature,[69] though obviously the detailed behavior depends on the specific dynamical rules. We remark that we did not measure rates of growth of various faces in our simulations — a subject of a possible future project — because this would require an algorithm for a dynamical, in-process identification of the crystal structure (for the program to decide where and which are the main growing cluster faces), which represents a nontrivial pattern recognition programming issue.

This mechanism seem to facilitate shape selection in the FCC case: Even for a relatively slow nonequilibrium growth, illustrated in Figure 9, a shape is obtained which involves only the (100) and (111) faces, and resembles the proportions of the equilibrium Wulff shape, also shown, that would be obtained for equal interfacial free energies of the two types of faces.

Let us now consider fast nonequilibrium growth for the FCC case. As illustrated in Figure 10, such growth involves nonlinear processes, as discussed earlier, and, as seen for other lattice symmetries, can yield well defined shapes for intermediate times, Figure 10a, with the leading growth mode involving vertexes advancing along (100) type directions. Less stable growth is possible for other parameter choices, e.g., Fig, 10b, including the situations where more than one unstable direction is observed, Figure 10c: (100) and (111), and yielding shapes ranging from distorted to ultimately chaotic.

### 3.5. *The HCP Lattice*

Recall that for the HCP lattice, we assumed that the nearest neighbor distance, form a central atom to its 12 neighbors, shown in Figure 11, is 1, in units of the free atom hopping steps, $l$. For this lattice, we estimated $\alpha_{cr} \approx 0.8$. We will use the Cartesian coordinates and "self-explanatory" notation for vector and plane orientation here as well, see Figure 11.

The minimal value of the surface energy, $6\varepsilon$, is achieved for atoms in the following locations (for lattice orientation shown in Figure 11): Two horizontal planes $(0\ 0\ 1)$, $(0\ 0\ -1)$. Twelve surfaces which are at $\theta \approx 62°$ with respect to the horizontal plane; each of these surfaces is actually corrugated (ridged); see contour *I* in Figure 11. Six vertical corrugated surfaces: see contour *V* in Figure 11. Figure 12 illustrates a relatively slow growth simulation, as well as the would be Wulff shape had the particle been in thermal equilibrium and with all these faces having equal interfacial free energies.

It is obvious that the faces identified, are not equivalent in dynamical growth. Indeed, a single atom attaching to (001), has binding energy $3\varepsilon$. The type-*I* faces have $4\varepsilon$, whereas the vertical faces (type-*V*) have $6\varepsilon$. For slow growth, $\tau_d < \tau_{layer}$, the cluster shape is not far form the Wulff shape, Figure 12, though *V*-faces are practically not formed, because of fast adsorption of free atoms on the "equator" of the particle, where the binding energy is large. Instead, wedges made of type-*I* faces are present.

For faster nonequilibrium growth, the cluster shapes can be quite different: see Figure 13. Here $\tau_d$ is initially small as compared to $\tau_{layer}$, and the shape is close to that of Figure 12. The shape then evolves as shown due to the dynamics of the spatially distributed diffusional flux, $\Gamma(x,y,z,t)$. The shape in Figure 13a, has its sharpest features close to the bases, at the edges of the faces (0 0 1) and (0 0 –1), where the "–1" is self-explanatory: recall that we are not using the Miller index notation. Therefore, diffusional flux is preferential in these regions. In addition, in a nonequilibrium regime the probability of an atom to move from type (001) faces to type *I*-faces, is larger than the opposite move (because the latter faces have stronger binding for single atoms). As a result, a prism is obtained; see Figure 13b. We again encounter the shape transformation mechanism found earlier for other lattices.

It is interesting to note that the growth is faster in the horizontal directions (Figure 13). In part this could be attributed to the packing of atoms along the *z* direction (Figure 11). However, the larger does the ratio $d_\perp / h$ get, the more of the growth asymmetry can be assigned to that the diffusional flux to the sides of the growing cluster becomes (somewhat) larger than that to the horizontal faces: $\Gamma_{side} > \Gamma_z$.

Furthermore, for larger times, see Figure 13c, the regions near the edges of the bases of the prism constitute preferential attachment locations. Apparently, in these regions $\tau_d > \tau_{layer}$, and therefore particle outflow along the surface in ineffective: The shape begins to bulge at the edges, not dissimilar to the late-time distortion of the vertexes of the cube in Figure 4c. However, in this case the evolution is more "orderly."

If $n_0$ is not too large (in real experiments this quantity is actually a decreasing function of time), then the on-surface drift of atoms between regions such a those marked by *a* and *b* in Figure 13c can prevent destabilization and yield growth of a hexagonal platelet shape. Otherwise, the protruding features such as *a* and *b* in Figure 13c, will become the bases for destabilization and formation of secondary structures. Figure 14 illustrates the latter type of growth. For the selected parameter values, the system is not controllable. However, the shape in Figure 14 did not yet become fully chaotic: Each of the protruding secondary-structure "blobs" can be loosely associated with one of the 18 corners in configurations such as Figures 12 or 13c.

## 4. Discussion and Additional Considerations

Our numerical simulations reported thus far, suggest the following key conclusions. Under certain specific conditions, straightforward processes that include arrival of singlet building blocks (our "atoms") to the growing cluster, detachment of these singlets, and also their drift on the cluster's surface, can account for the formation of particle shapes defined by a set of the densest-packed, low-index crystal-symmetry faces.

The most important condition is that *no large-scale defects are present*, that would dominate the singlet kinetics and the overall rates of the growth dynamics. Interestingly, colloids, given their generally polycrystalline nature, developed by aggregation of nanocrystalline singlets, or by other internal restructuring processes that are presently not well studied or understood, satisfy this condition. The constituent nanocrystalline subunits are apparently not correlated in their crystal-structure orientation or defect continuity: To the extent evidence is available, the subunits are likely separated by amorphous interlayers. Of course, the actual mechanism of the singlet attachment, surface dynamics, overall cluster compactification, and ultimately singlets getting "cemented in place" to become part of the growing structure, buried by later arriving singlets, should not be expected to bare any resemblance to our simple model of forced ideal crystal formation. However, the net result is apparently the same, as suggested by ample experimental evidence.

Another condition has been the use of dynamical rules that mimic thermal, free-energy driven relaxation. Colloid synthesis generally involves processes far from equilibrium or steady-state dynamics. However, energy-related transition rates of the type defined in Section 2, are not that far fetched for attachment of larger than atomic-size singlets, as long as the explicit temperature dependence used in Boltzmann-form weights, is not taken literally.

The applicability of our observations for nanocrystals is also conditioned on that they have no dynamics-controlling, global defect structures, and should probably be verified on a case by case basis.

The densest-packed, low-index crystal-symmetry faces, some of which dominate the generated crystal shapes, also figure in the thermal-equilibrium Wulff construction in typical situations. However, our nonequilibrium-growth particles assume similar proportions at best only when the growth is slow (or when we consider the steady-state growth as defined in Section 2.3).

For faster growth, some of the faces are no present, the particle proportions are different, and we have identified some "rule of thumb" regularities connected to the binding energies of atoms within a flat surface and those attaching as singlets on such a filled surface. Some of the observed regularities have been associated with the distribution of the diffusional fluxes to sharp surface features. In some situations direct identification is possible of sharp surface features which are expected to cause, and do lead to, distortions and instabilities, and serve as regions on top of which new structures develop.

It would be interesting to extend our simulations to study formation of other, larger-time shapes, as well as other particle-growth processes. However, as pointed out in Section 2, the simulations were "large-scale:" resource and CPU time consuming. Therefore, we leave such explorations to future work. Furthermore, the applicability of the present model to larger-scale growth has not been established.

We did run some simulations whereby an attempt was made to introduce protruding features to control the cluster distortion. For an illustration, let us point out that in many experimental works colloid particles shaped as ellipsoids of revolution were synthesized. Therefore, we attempted a SC simulation for which the initial seed was ellipsoidal, with the large axis made 1.5 times longer than the short axis. While a guess can be made that the two large-axis ends will serve as the starting regions for fast growth, in particular, collecting larger diffusional flux, it turns out to be an oversimplification: In addition to elongation, secondary distortions develop, see Figure 15, along directions typical for the SC symmetry. We attribute this to the formation of fragments of the (111) type faces due to fluctuations. These regions have large atom-binding energies and therefore serve of "seed regions" for faster growth and, ultimately, possible instability.

We know experimentally that in many systems, growth of long rods and other structures of uneven proportions is quite common. Can our approach reproduce such large-time regimes? The simulations would be formidable and beyond our present numerical capabilities. However, we did find one manageable (by a lengthy run) example, shown in Figure 16.

For the BCC lattice, Figure 16, the rhombic dodecahedron shape (Figure 8) is initially symmetric with respect to the $x$, $y$, $z$ directions. The distortion along the $z$ axis for later times, Figure 16, is a result of random fluctuations. Specifically, the tendency to elongate is already seen in Figure 8e, in which $l_x = 75$, $l_z = 69$, so that there is preferential growth in the $x$ direction. The rod-like distortion seems to grow without developing instabilities, though we did not attempt to follow the dynamics to the time scales of a possible long rod formation (due to computational resource limitations). Furthermore, the reported result is not fully reproducible: This growth regime is too close to "chaotic." The emergence of a rod-like shape did not happen for a fraction of the runs. Instead, for some runs clusters ranging from even-shaped to chaotic were obtained, starting from those shown in Figure 8, with details dependent on the model parameters.

In summary, we hope that our present numerical results will somewhat demystify the long-standing open problem of the origin of shape selection in colloid and nanoparticle synthesis and shed light on the mechanisms and conditions for obtaining particle shapes sought after for applications. The status of this research field is still far from predictive, and additional theoretical studies, as well as confrontation of the modeling results with experimental data, are needed. Our results suggest an emphasis on experimental probes of the morphological features on the scales of the particle as a whole, to test the conclusion/conjecture that absence of persistent defect structures, which could globally influence the face growth dynamics, is crucial to shape-selection in fine-particle synthesis.

The authors gratefully acknowledge instructive discussions and collaboration with G. P. Berman, D. V. Goia, I. Halaciuga, S. Libert, E. Matijević, and I. Sevonkaev, as well as research funding by the U.S. National Science Foundation (grant DMR-0509104).

**FIGURES**

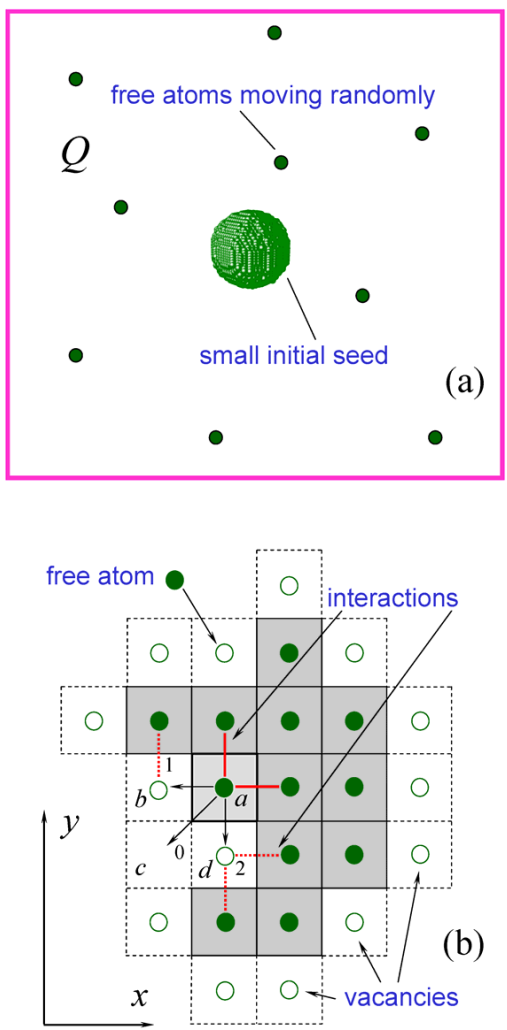

**Figure 1.** (a) System geometry: the initial seed is placed in the three-dimensional region $Q$. (b) An illustration of the cluster growth dynamics, here for simplicity shown in two dimensions. The attached cluster atoms are exactly "registered" with the lattice structure of the cluster, defined starting from the initial seed (which is not specially marked): the centers of the shaded squares. Each atom interacts with nearest neighbors, here along the lattice directions of the type (10). Thus, the maximum of interactions is 4 for our square-lattice illustration, and each interaction is assigned energy $\varepsilon < 0$; see text. The numbers in the cells labeled $b, c, d$, enumerate new interactions for the atom presently in cell $a$, if it hops to the respective location ($b$, $c$, or $d$). Free atoms are captured into the surface vacancies which are nearest-neighbor to at least one of the cluster atoms, marked by open circles, once they hop into the center part of a vacant lattice-cell, at a distance less than 1/2 hopping-lengths to that cell's center. The captured atom is then instantaneously "registered" with the lattice by positioning exactly at that cell's center.

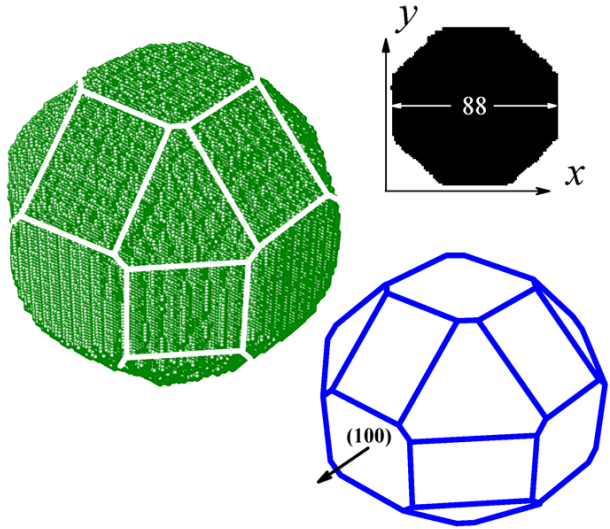

**Figure 2.** Steady state SC lattice simulations for variant *A* of the displacements/detachments for surface atoms. The simulation details and parameter values are given in Section 3.1. The resulting particle shape is shown for the cluster of $3.8 \times 10^5$ atoms, which was in steady state with a dilute solution of $2.4 \times 10^3$ free atoms. Also shown is the projection of the cluster shape onto the *xy* plane, as well as a shape formed by lattice planes of the types (100), (110), (111) by an equilibrium Wulff construction (assuming that they all have equal interfacial free energies).

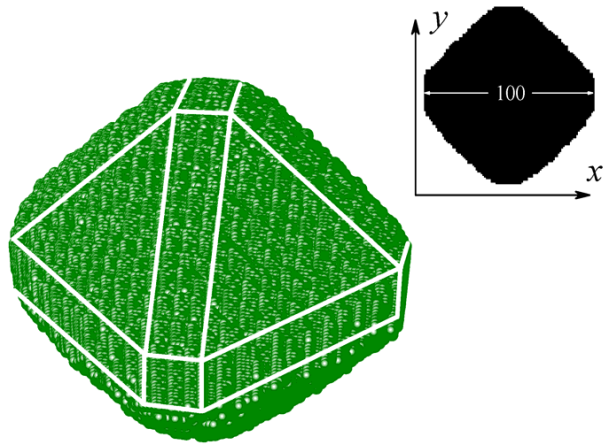

**Figure 3.** Steady state SC lattice simulations for variant *B* of the displacements/detachments for surface atoms. The dynamical rules were the same as for variant *A*, presented in Figure 2 (and detailed in Section 3.1). Also shown is the projection of the cluster shape onto the *xy* plane. The dilute solution contained $1.3 \times 10^3$ free atoms in this case.

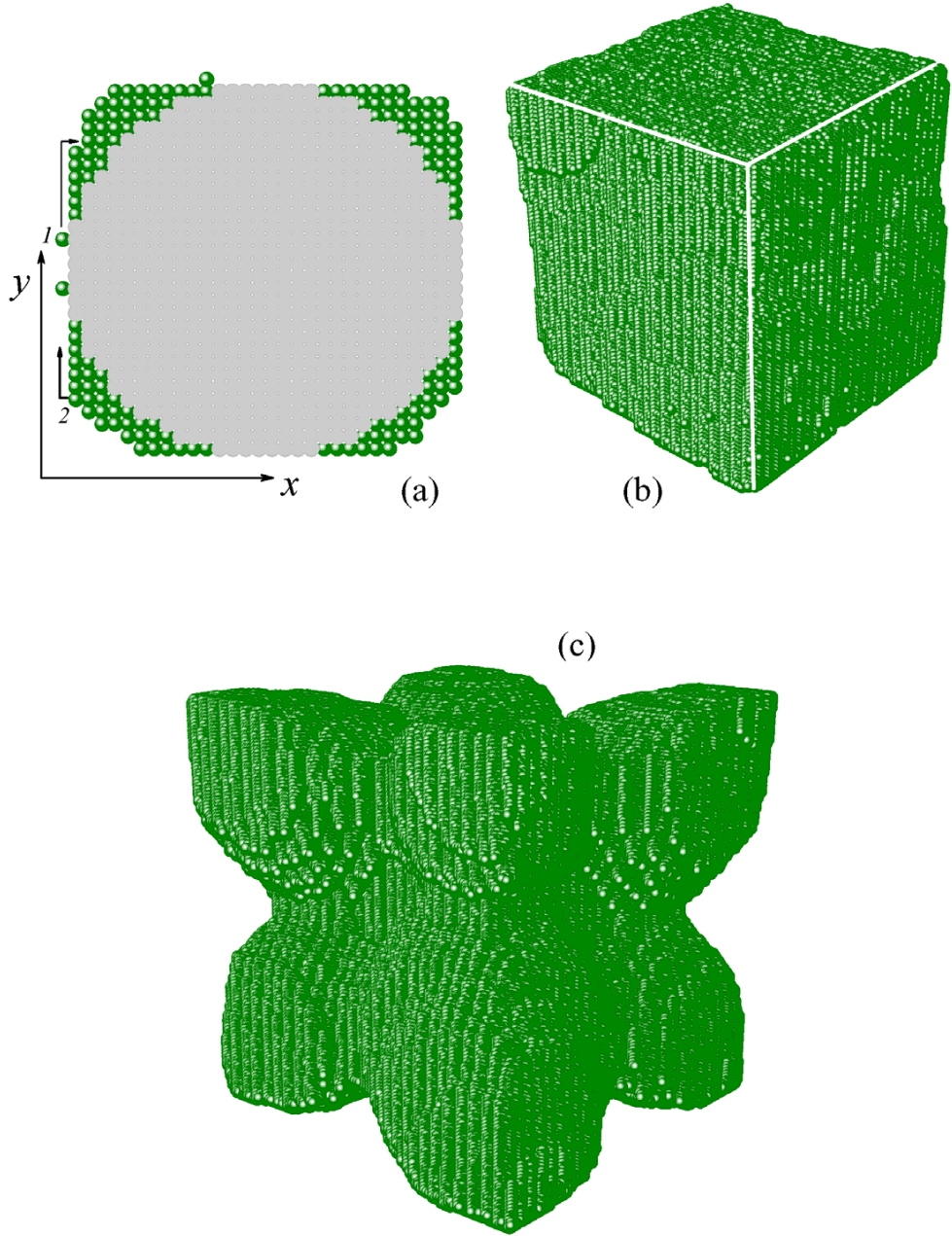

**Figure 4.** Nonequilibrium SC-lattice cluster shapes in variant $A$ (detailed in Section 3.1). Here $n_0 = 1.8 \times 10^{-3}$, and the radius of the initial seed $r_0 = 15$. (a) A projection onto the $xy$ plane, of the initial atoms in the seed (gray circles) and in the growing cluster (green circles), with parameter values $p_0 = 0.6$, $\alpha = 2.5$, shown at $t = 3 \times 10^5$, illustrating the emergence of the cubic shape. (b) The same cluster at a later time, $t = 2.5 \times 10^6$, containing $4.5 \times 10^5$ atoms, with the cube edge length 77. (c) Cluster grown with different parameter values, $p_0 = 0.8$, $\alpha = 3.5$, shown at time $t = 5.2 \times 10^6$, containing $1.8 \times 10^6$ atoms, and of characteristic size 125.

– 28 –

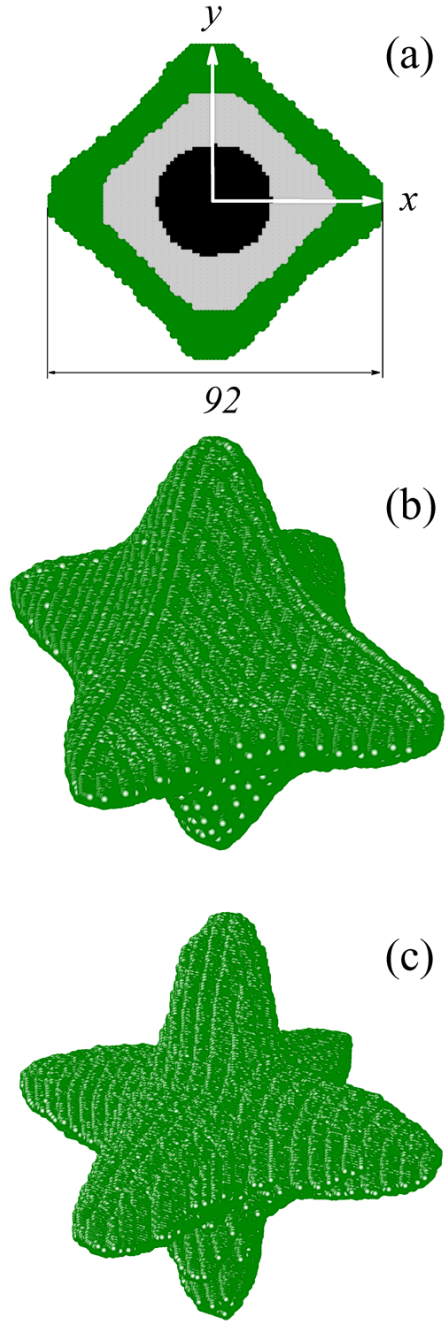

**Figure 5.** Nonequilibrium SC cluster shapes in variant $B$. Here $p_0 = 0.6$, $\alpha = 2.5$, $n_0 = 1.8 \times 10^{-3}$, and the radius of the initial seed $r_0 = 15$. (a) A projection onto the $xy$ plane, for three different times: $t = 10^5$, $2 \times 10^6$, $3 \times 10^6$. (b) The shape found for time $t = 7 \times 10^6$, with $1.18 \times 10^6$ atoms in the cluster, which can be inscribed in a sphere of diameter 210. (c) The shape for larger availability of free atoms: $n_0 = 3 \times 10^{-3}$, with $3.8 \times 10^5$ atoms in the cluster ($t = 2.4 \times 10^6$), which can be inscribed in a sphere of diameter 175.

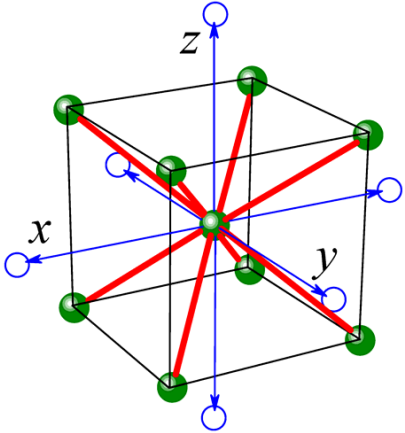

**Figure 6.** Eight nearest neighbors and six next-nearest neighbors of an atom in the BCC lattice.

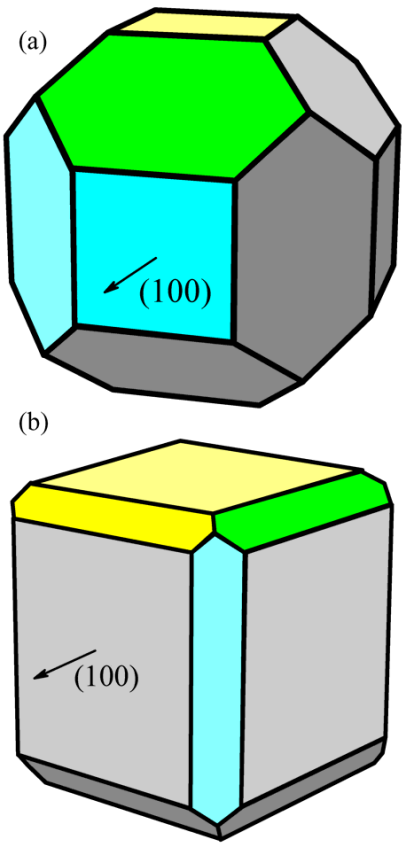

**Figure 7.** (a) Wulff construction based on the assumption of dominance of the (100) and (110) type faces, with equal interfacial free energies, for the BCC lattice. (b) The actual cluster shape obtained in the steady-state regime simulation, with $p_0 = 0.7$, $\alpha = 2$, and $R(t=0) = 45$.

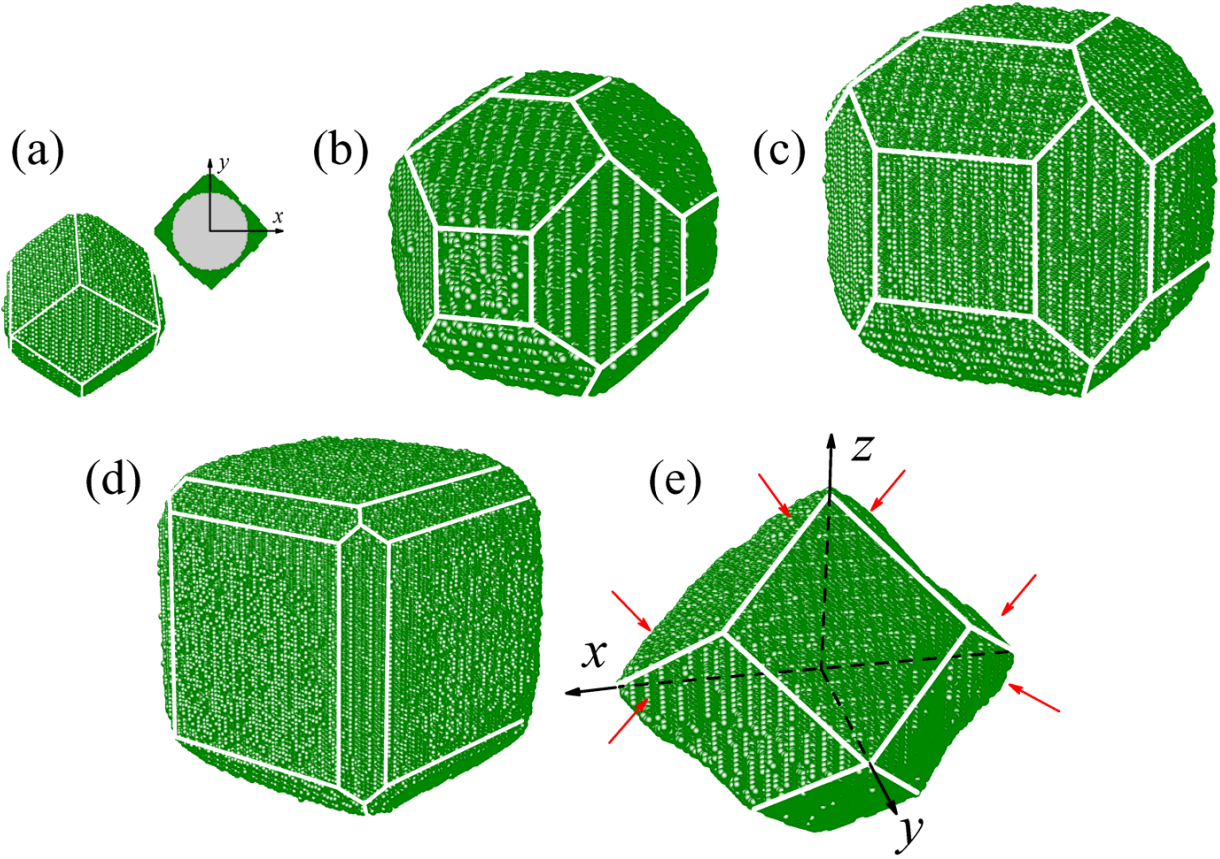

**Figure 8.** Nonequilibrium regime for the BCC lattice. (a) Parameter values $r_0 = 15$, $\alpha = 2$, $p_0 = 0.6$, $n_0 = 1.8 \times 10^{-3}$: Formation of a rhombic dodecahedron from the type-(110) planes at the initial growth stages, with the caliper (maximal) cluster dimension equal 42. The inset shows the $xy$ plane projection of the initial seed (gray) and the formed cluster. (b) and (c) Evolution of the cluster shape at later times, for cluster sizes 59 and 68, respectively, measured along the $x, y, z$ directions. The time sequence for (a), (b), (c) was $t = 5 \times 10^5, 2 \times 10^6, 3 \times 10^6$. (d) Parameter values changed to $p_0 = 0.7$, $n_0 = 2.25 \times 10^{-4}$: Cluster shape when grown to size 68 ($t = 3 \times 10^7$). (e) Parameter values changed to $p_0 = 0.6$, $n_0 = 4.5 \times 10^{-3}$: rhombic dodecahedron of size 69 ($t = 10^6$). The arrows point to regions in which one can see the local piling up of the attached atoms (an onset of destabilization).

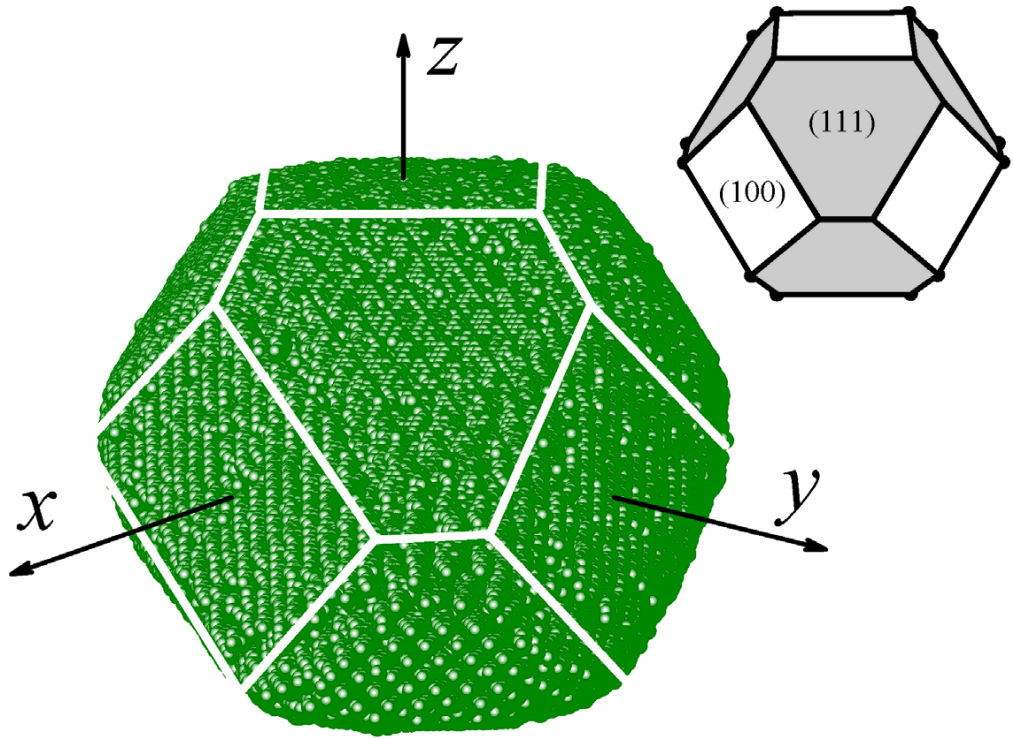

**Figure 9.** A relatively slow-growth nonequilibrium FCC-lattice simulation: $\alpha = 1$, $p_0 = 0.7$, $n_0 = 2.25 \times 10^{-4}$. The particle dimensions along the coordinate axes are 127 (63 lattice constants), and the total number of atoms in it is $6.82 \times 10^5$. The run time was $t = 2.2 \times 10^7$. The inset shows the Wulff-construction shape based on the (100) and (111) type faces, assuming equal interfacial free energies for both.

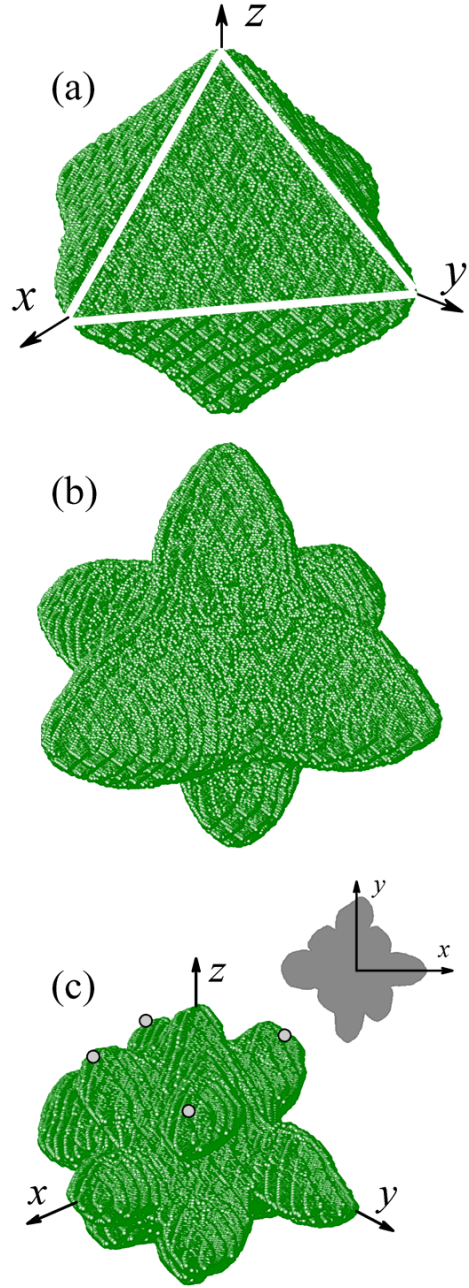

**Figure 10.** Nonequilibrium FCC-lattice simulation: (a) Parameter values $\alpha=1$, $p_0=0.7$, $n_0=1.8\times 10^{-3}$. The dimensions along the coordinate axes are 198, and the total number of atoms is $7\times 10^5$ ($t=3.6\times 10^6$). (b) Parameter values $\alpha=1.4$, $p_0=0.7$, $n_0=1.8\times 10^{-3}$. The dimensions along the coordinate axes are 233, and the total number of atoms is $1\times 10^6$ ($t=3.6\times 10^6$). (c) Parameter values $\alpha=3.5$, $p_0=0.8$, $n_0=9\times 10^{-4}$. The dimensions along the coordinate axes are 214, and the total number of atoms is $1\times 10^6$ ($t=7\times 10^6$). The gray circles denote protrusions grown in the (111) directions. The inset gives the projection of this shape onto the $xy$ plane.

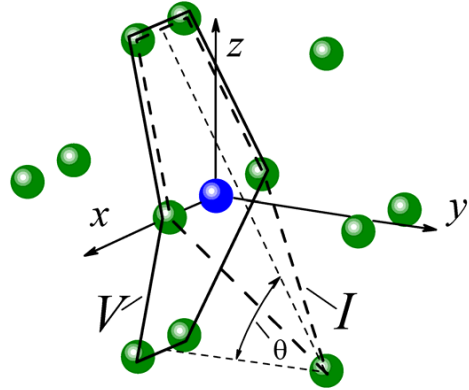

**Figure 11.** Features of the HCP lattice: The structure is shown assuming that the straight chains of atoms are positioned along the *x*-axis direction. The central atom (blue) interacts with 12 nearest neighbors (shown in green). The contours $V$ and $I$ denote corrugated surfaces for which $\bar{\varepsilon}_{surf} = -6\varepsilon$ is minimal (see text).

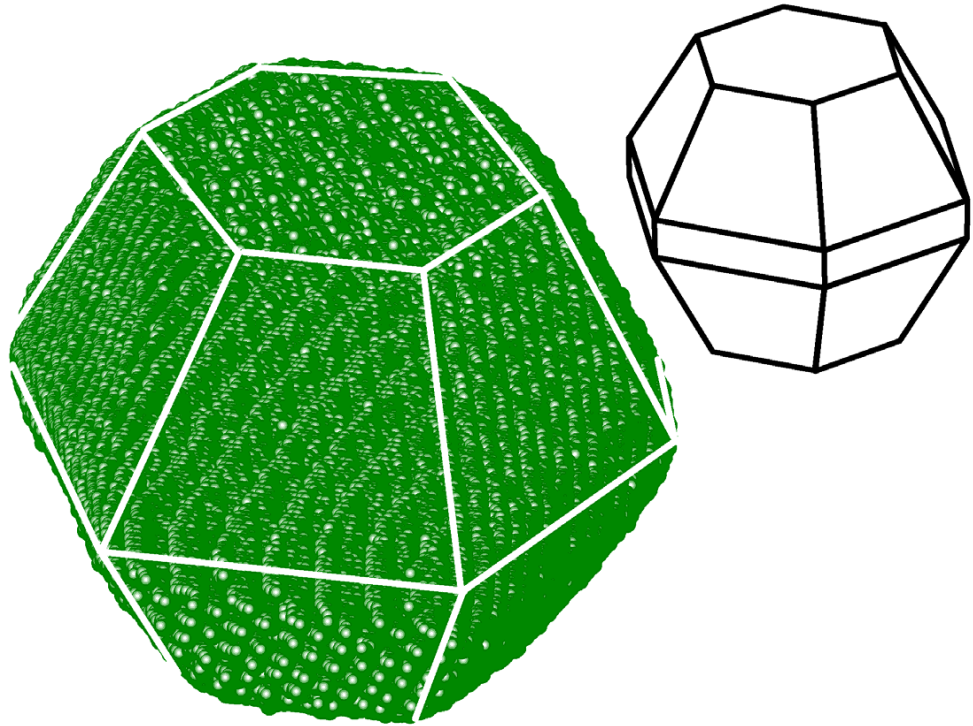

**Figure 12.** The HCP lattice simulation for slow growth, with parameter values $p_0 = 0.7$, $\alpha = 0.9$, $n_0 = 6 \times 10^{-4}$. The cluster contains $1.122 \times 10^6$ atoms ($t = 4.7 \times 10^6$). The inset shows the Wulff shape based on all the faces with strong binding (see text) which would be obtained under the condition of equal interfacial free energies for all of them.

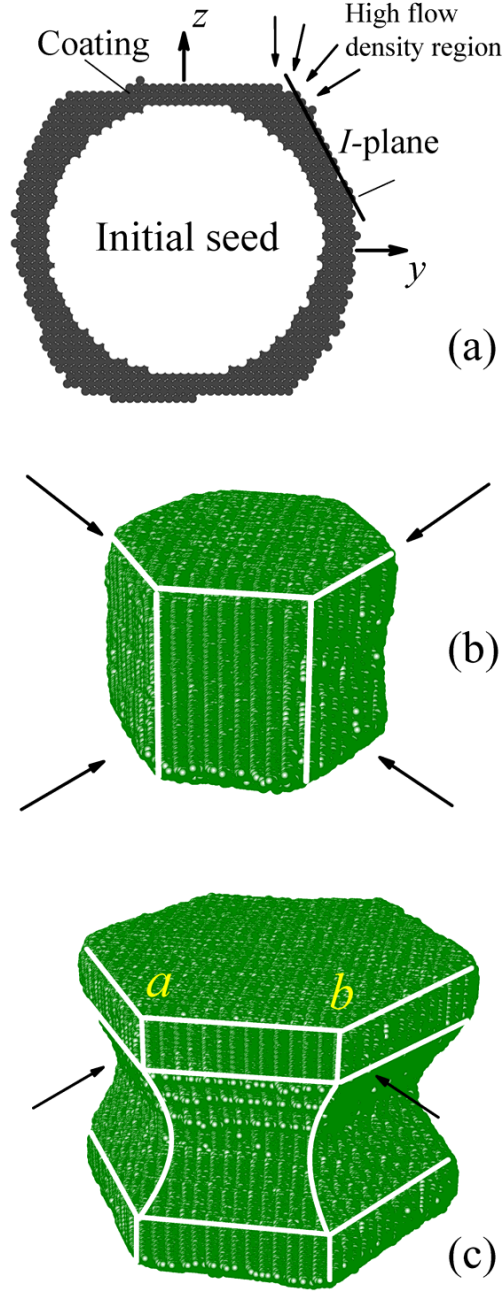

**Figure 13.** The HCP lattice simulation for faster growth, with parameter values $p_0 = 0.6$, $\alpha = 0.9$, $n_0 = 1.8 \times 10^{-3}$. (a) The initial stage of the growth: Shown is the projection of the shape into the $yz$ plane ($t = 5 \times 10^5$). (Only the atoms added to the initial seed as a "coating" are shown.) (b) Formation of a right hexagonal prism, of height $h = 66$, with the base of size (diameter of the circumscribed circle) $d_\perp = 83$, shape ratio $d_\perp / h = 1.26$, and with $3.77 \times 10^5$ atoms in the particle ($t = 3 \times 10^6$). (c) Cluster shape at a later time: $h = 96$, $d_\perp = 134$, shape ratio $d_\perp / h = 1.4$, and $1.271 \times 10^6$ atoms ($t = 6.5 \times 10^6$).

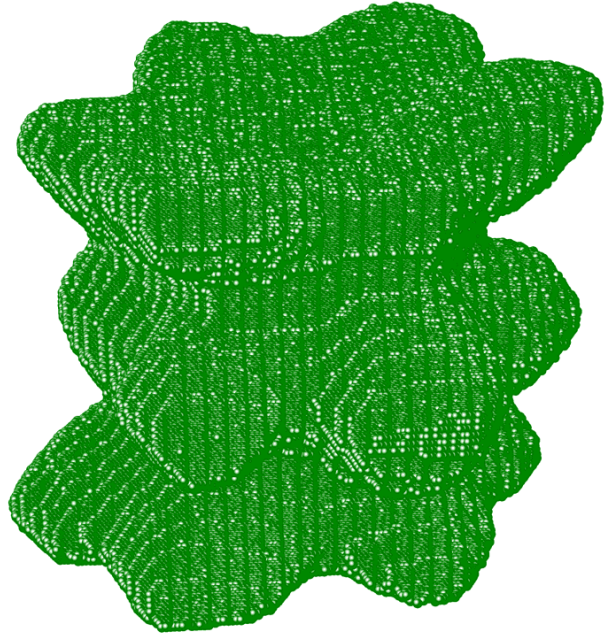

**Figure 14.** Onset of a multimode unstable growth: The HCP lattice simulation with $p_0 = 0.5$, $\alpha = 1$, $n_0 = 1.8 \times 10^{-3}$. Here the cluster contains $1.278 \times 10^6$ atoms ($t = 6 \times 10^6$).

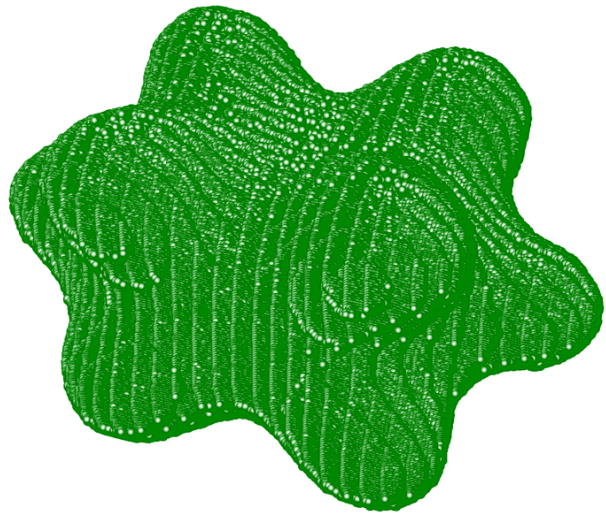

**Figure 15.** The SC lattice simulation (type-$A$ dynamics) for an initially elongated (in the horizontal direction) seed. Here $p_0 = 0.5$, $\alpha = 2.3$, $n_0 = 1.8 \times 10^{-3}$, $t = 8 \times 10^6$, and the cluster contains $1.9 \times 10^6$ atoms.

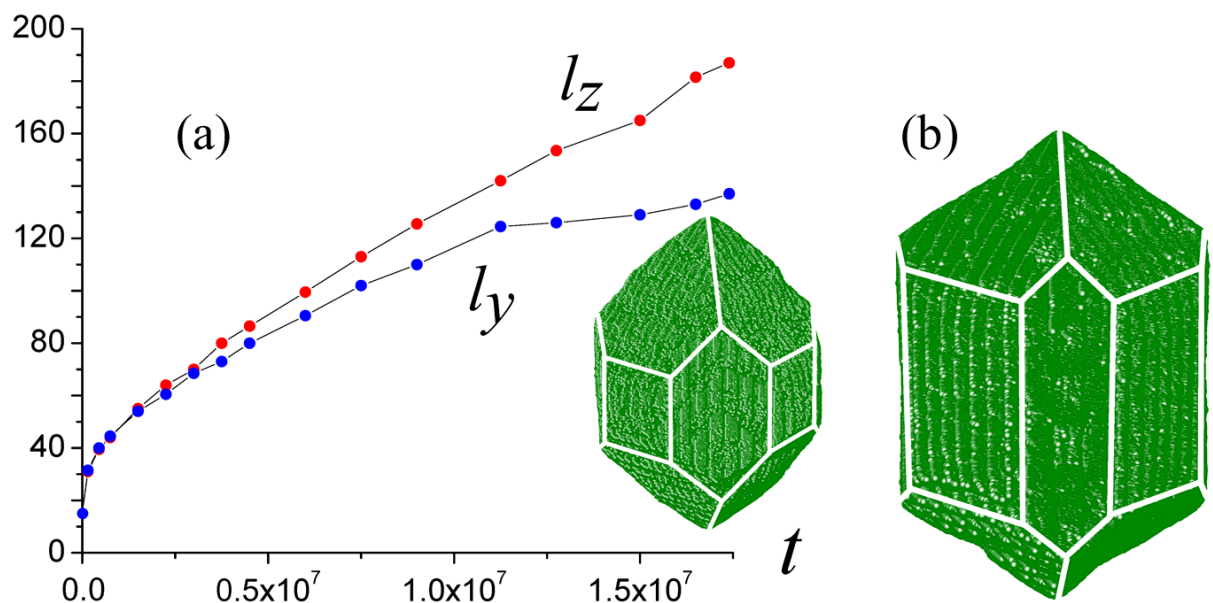

**Figure 16.** Onset of a rod-like shape growth in a long-run nonequilibrium BCC lattice simulation. Here $p_0 = 0.6$, $\alpha = 2$, $n_0 = 1.125 \times 10^{-3}$. (a) The time-dependence of the transverse sizes, $l_y (\approx l_x)$, and of the longitudinal, $l_z$, size. The inset shows the shape of the cluster for $t = 1.75 \times 10^7$. (b) $t = 5.2 \times 10^7$, at which time the cluster contained $\sim 5 \times 10^6$ atoms, $l_z = 240$, $l_x = 140$.